\documentclass{elsarticle}
\usepackage{graphics}
\usepackage{amssymb}
\usepackage{amsmath}
\usepackage{wasysym}
\usepackage{latexsym}
\usepackage{graphicx}
\usepackage{epsfig}
\usepackage{subfigure}
\usepackage{float}
\usepackage{color}
\usepackage{lineno}
\usepackage{verbatim}
\usepackage{enumitem}
\usepackage[T1]{fontenc}

\journal{Physica A: Statistical Mechanics and its Applications}
\begin{document}

\begin{frontmatter}

\title{Competitive Social Mobilization in Threshold Models of Collective Action}

\author[mymainaddress]{Bianca Y.  S.  Ishikawa}
\ead{yumibianca@usp.br}

\author[mymainaddress]{Jos\'e F.  Fontanari\corref{mycorrespondingauthor}}
\cortext[mycorrespondingauthor]{Corresponding author}
\ead{fontanari@ifsc.usp.br}

\address[mymainaddress]{Instituto de F\'{\i}sica de S\~ao Carlos, Universidade de S\~ao Paulo, 13566-590 S\~ao Carlos, S\~ao Paulo, Brazil}

\begin{abstract}
Social mobilization often fails not for a lack of collective interest, but because of fierce competition between rival movements for the same limited pool of participants. We generalize the classic threshold model of collective behavior to analyze this competitive aggregation,  exploring how populations with diverse participation thresholds navigate multiple, mutually exclusive causes. Focusing on the conditions necessary for a single consensus movement to encompass an entire population, our analysis reveals that the outcome of social competition depends critically on the stability of individual dispositions. In  quenched environments where participation thresholds are fixed, increasing resistance initially allows a dominant movement to suppress its competitors; however, further resistance triggers a sudden collapse into total fragmentation as low-threshold instigators become too rare to sustain growth. Conversely, in  annealed  environments where opinions are fluid, higher resistance paradoxically drives a  winner-takes-all  consensus. In this fluid scenario, massive movements can only be avoided through a deliberate  divide-and-conquer  strategy. In both cases, the transitions between pulverized and massive movements are discontinuous. These findings demonstrate that the effectiveness of social control depends entirely on environmental stability: raising the cost of participation can either forge unity or shatter collective action into insignificance.
\end{abstract}


\end{frontmatter}

\section{Introduction}\label{sec:intro}

The collective dynamics of social systems are fundamentally shaped by how individuals navigate choices among multiple, competing alternatives \cite{Coleman_1964,Gavrilets_2015}. Whether observing the spontaneous formation of casual groups in public settings \cite{Coleman_1961, White_1962, Fontanari_2023}, the emergence of dominant technologies in a marketplace \cite{Katz_1985, Arthur_1989}, or the mobilization of protesters across various political factions \cite{McCarthy_1977, McAdam_2001, Bakke_2012}, the transition from individual preference to macroscopic order often involves a process of competitive aggregation.

While classical sociological models have long utilized the concept of  ``thresholds" to describe binary decisions---such as whether to move or stay in a neighborhood \cite{Schelling_1971}, or whether  to join a single riot or strike  \cite{Granovetter_1978}---real-world conflicts and social movements are rarely characterized by such cohesion.  
Instead, as seen in complex civil wars or modern political unrest, a population is often fragmented into numerous organizations and movements simultaneously vying for the same pool of participants. This fragmentation is the structural signature of competitive aggregation: a dynamical process where the growth of one collective entity necessarily limits the potential of its rivals \cite{Hannan_1977}.

In this work, we extend Granovetter's threshold model of collective behavior \cite{Granovetter_1978}---originally designed for binary decisions---to accommodate choices among multiple, competing movements. In our framework, an individual decides to join a specific movement only when the number of existing participants reaches or exceeds their unique participation threshold, thereby instigating a competitive dynamic between rival entities.

This approach synthesizes the threshold concept with the casual group formation framework pioneered in the 1960s \cite{Coleman_1961, White_1962, Fontanari_2023}, which has recently been adapted to explicit spatial settings and face-to-face interaction networks \cite{Starnini_2013, Starnini_2016, Mariano_2025}. However, our model diverges from traditional casual group formulations in one critical respect: membership is permanent. This once-and-for-all decision logic reflects a social environment defined by intense peer pressure or mutual distrust, where crossing between groups is prohibited. This observed persistence, coupled with individual threshold heterogeneity, stands in sharp ontological contrast to recent models of protest dynamics that treat collective action as a memoryless stochastic process. Such models often assume that individuals are fluidly interchangeable, entering or exiting a single group at rates largely decoupled from personal history or fixed internal dispositions \cite{Petrovskii_2020, Alsulami_2022, Volpert_2024, Volpert_2025}. By contrast, our results suggest that the stickiness of collective identity is fundamentally anchored in the very individual-level variances that these mean-field approaches tend to smooth over. Notably, while we depart from modern fluid-membership models, our assumption of permanent commitment remains strictly in accord with Granovetter’s original formulation \cite{Granovetter_1978}.

Our primary objective  is to determine the macroscopic outcome of this multi-group competition: specifically, whether a  consensus group that encompasses the vast  majority of the population emerge, or whether the social fabric remains pulverized  into small, fragmented groups. Starting from an initial state where all individuals are isolated, we investigate how the interplay between fierce inter-group competition and individual decision-making drives the scale of collective action.

The resolution of this question depends fundamentally on the distribution of participation thresholds $T$ across a population of size $N$ \cite{Granovetter_1978}. To explore this, we employ a cumulative threshold distribution $P(T \leq k) = (k/N)^\gamma$ for $k = 1, \ldots, N$, where the aggregation exponent $\gamma \geq 0$ allows us to interpolate between distinct sociological regimes.

When $\gamma = 1$, the model yields a uniform distribution of thresholds, representing a linearly heterogeneous population. In the limit $\gamma \to 0$, we recover a  casual group dynamics scenario (absent disaggregation), where individuals join any group regardless of its size. In this regime, the social barrier to entry vanishes and every individual acts as a potential agitator, willing to embark on any available movement. Conversely, large values of $\gamma$ describe a population of highly demanding or risk-averse individuals. In this case, the distribution is heavily skewed toward high thresholds, implying that individuals will only join large-scale groups where the perceived costs or risks of participation are sufficiently diluted among a vast number of participants.

While our choice of threshold distribution depends explicitly on the population size $N$---a global parameter that defines the upper bound of collective action---this dependency does not limit the model's generalizability. Because our primary focus is the thermodynamic limit $N \to \infty$, the scale-dependence is effectively mitigated, allowing us to characterize universal behaviors in the emergence of social groups.

Our results demonstrate that the temporal nature of the disorder—whether thresholds are fixed at the outset (quenched) or renewed at each step (annealed)—diametrically opposes the dynamical consequences of the exponent $\gamma$. In the thermodynamic limit, the quenched scenario with $\gamma \ge \gamma_c^{(q)} \approx 1$ leads to a completely fragmented state; the convex distribution of thresholds creates a population of ``stubborn" individuals whose high resistance blocks group initiation. We find that a discontinuous transition occurs at $\gamma_c^{(q)}$, separating a regime with macroscopic groups ($\gamma < \gamma_c^{(q)}$) from one where only isolates have a macroscopic representation ($\gamma \ge \gamma_c^{(q)}$). Remarkably, a consensus group emerges in this scenario only as $\gamma$ approaches $\gamma_c^{(q)}$from below. In these quenched configurations, the dynamics reach an absorbing state where all remaining isolates possess thresholds that exceed the size of the largest available group.

Conversely, in the annealed scenario, the renewal of thresholds transforms $\gamma$ into a dynamic attachment preference. Here, as the population size $N \to \infty$, a value of $\gamma > \gamma_c^{(a)} \approx 1.5$ triggers a  ``winner-takes-all" dynamic, where super-linear attachment enforces a total consensus state. In this limit, the discontinuous transition at $\gamma_c^{(a)} $ separates a  ``pulverized" regime ($\gamma \leq \gamma_c^{(a)} $) from a state where a single macroscopic group encompasses the entire population, leaving no isolates in the final absorbing configuration. Consequently, high values of $\gamma$ prevent collective action in the quenched case by stifling initiation, yet facilitate it in the annealed case by suppressing competition. These findings highlight how the temporal stability of individual dispositions fundamentally reshapes the mechanisms of social aggregation in large-scale systems.

The remainder of this paper is organized as follows.  In Section \ref{sec:mod}, we formally define the threshold model and the governing dynamics of group formation. Section \ref{sec:qq} examines the quenched scenario; we argue that this case is the more realistic representation of human social behavior, as it assumes individual dispositions are stable over the timescales of collective action. However, the quenched case presents significant mathematical challenges, leading us to introduce the annealed scenario in Section \ref{sec:aa}. While physically distinct, the annealed scheme serves as a powerful analytical proxy that allows for an analytical  formulation of the problem in the thermodynamic limit $N \to \infty$. This approach yields exact solutions for the boundary cases of $\gamma = 0$ and $\gamma = 1$, the derivations of which are provided in   \ref{app:A} and  \ref{app:B}, respectively. Finally, Section \ref{sec:disc} offers a comparative discussion of our findings and their implications for social tipping points.

\section{The Threshold Model}\label{sec:mod}

We consider a population of $N$ agents, initially in a state of total fragmentation where each agent forms a group of size one (an isolate).  In the following technical sections, we adopt the term  agent rather than  individual to emphasize the computational and algorithmic framework of our approach.  The formation of groups is driven by agent preferences regarding group size. The decision of each agent $i$ to join a group is determined by its threshold $T_i$, representing the minimum group size it is willing to join \cite{Granovetter_1978}.

The thresholds $T_i$ are independent, identically distributed (i.i.d.) random variables defined by the discrete probability mass function 
\begin{equation}\label{PT}
P(T_i = k) = \frac{1}{N^\gamma} \left[ k^\gamma - (k-1)^\gamma \right],
\end{equation}
where $k \in \{1, \ldots, N\}$. The cumulative distribution function of this threshold mechanism is exactly $P(T_i \le k)=(k/N)^\gamma$. To generate random variates satisfying this distribution, we first define a continuous random variable $x$ on the interval $x \in [0, 1]$ whose cumulative distribution function $F(x) = x^\gamma$ yields the probability density $f(x) = \gamma x^{\gamma-1}$. The discrete threshold $T_i$ is then set via $T_i = 1 + \lfloor N x \rfloor$. Numerically, $x$ is obtained efficiently using the inverse transform method as $x = u^{1/\gamma}$, with $u$ being drawn from a uniform distribution $U[0, 1]$.

When $\gamma = 1$, Eq. (\ref{PT}) simplifies to $P(T_i = k) =  [k^1 - (k-1)^1] /N= 1/N$,  so the thresholds $T_i$ are uniformly distributed across the discrete interval $k \in \{1, \ldots, N\}$. In the limit $\gamma \to 0$, the distribution converges to a threshold distribution concentrated entirely at $T_i=1$, implying that any group satisfies the aggregation requirement. This effectively nullifies the selective role of the threshold mechanism, causing the aggregation dynamics to be determined purely by chance. Consequently, this limit corresponds precisely to the original model of casual group dynamics where an isolated agent joins any other group with uniform probability. Hence, $\gamma$ plays the role of an attachment exponent, tuning the dynamics from random aggregation ($\gamma=0$) to strong preferential attachment to large groups ($\gamma > 1$).

The dynamics proceed via a Monte Carlo simulation in discrete time steps. We define the unit of time (one Monte Carlo step) as $N$ elementary steps, where in each step an agent  is selected at random from the population for a potential update. The fundamental aggregation rules are defined as follows: at each elementary  step, an agent $i$ is selected at random from the population. If agent $i$ is already a member of a group (size $>1$), no action is taken. If $i$ is an isolate, it attempts to join a group by selecting a target group $j$ uniformly at random from the pool of all existing groups. The aggregation event is accepted if and only if the size of the target group, $S_j$, satisfies the agent's threshold condition
\begin{equation}
S_j \ge T_i.
\end{equation}
If this condition is met, the agent joins group $j$, increasing its size by one ($S_j \to S_j + 1$). Otherwise, the attempt is rejected, and the agent remains isolated. 

It is important to emphasize a fundamental difference between the dynamics proposed here and the original threshold model introduced by Granovetter \cite{Granovetter_1978}. In Granovetter's formulation,  agents face a binary choice---to participate in a collective action or not---based on the total number of others who have already done so. Conceptually, this corresponds to a system with a single active group growing amidst a pool of inactive agents. In contrast, our model describes a scenario of  competitive aggregation,  where multiple groups can nucleate, coexist, and grow simultaneously. An isolated agent is not restricted to joining a unique `main' group but encounters various groups at random and may join any of them provided the size condition is met. This generalization allows for the emergence of complex structural configurations, such as the fragmentation of the population into multiple rival factions versus the formation of a single unified consensus, outcomes that are precluded by design in the single-group formulation.

In the following sections, we distinguish between two scenarios: the more realistic \textit{quenched} scenario, where the threshold $T_i$ assigned to agent $i$ at $t=0$ remains fixed throughout the simulation, and the \textit{annealed} scenario, where a freshly generated threshold $T_i$ is assigned to agent $i$ at each update attempt. In both scenarios, the dynamics eventually freeze (albeit for different reasons) into absorbing configurations. We focus on characterizing these final states, which are fully described by the cluster size distribution $n_k$, defined as the number of groups of size $k$ (for $k=1, \ldots, N$). The constraint of fixed population size implies that $\sum_k k n_k = N$. The quantities of interest are the number of isolated agents $n_1$, the total number of groups $M=\sum_k n_k$, and, most importantly, the number of agents in the largest group,  $S_{max}$.  We typically perform $10^4$ independent simulations and present the properly normalized average results, namely, $\phi_\infty = \langle n_1 \rangle /N$, $\mu_\infty = \langle M\rangle/N$, and $\rho_\infty = \langle S_{max}\rangle/N$. Here, the brackets $\langle \cdot \rangle$ denote the average over independent runs, and the subscript $\infty$ indicates that measurements are taken in the asymptotic time limit, once the dynamics have frozen.

\section{The Quenched Scenario}\label{sec:qq}

In this scenario, the thresholds $T_i$ are fixed at the start (quenched disorder) and remain unchanged throughout the simulation. Consequently, the cluster size counts $n_k$ act as the fast dynamical variables, evolving against the background of frozen thresholds.

For computational efficiency, we implement an optimized version of the general rejection-based dynamics described earlier. Instead of performing random trials that result in frequent rejections, an isolated agent $i$ scans the system to identify the subset of available groups---those that satisfy $S_k \ge T_i$. If this subset is non-empty, the agent selects one group uniformly at random from the available set and joins it immediately. If the subset is empty, the agent remains isolated. In the quenched scenario, where thresholds are intrinsic to the agents, this global search implementation generates a sequence of successful aggregation events statistically identical to the original rejection-based dynamics, differing only by a rescaling of the physical time.

The simulation iterates until the system reaches a frozen (absorbing) configuration. This occurs when the thresholds of all remaining isolates are strictly larger than the size of the largest group in the system ($S_{max}$). At this point ($T_i > S_{max}$ for all remaining isolates $i$), no further aggregation events are possible, and the distribution of group sizes becomes static.

\begin{figure}[th] 
\center
 \includegraphics[width=1\columnwidth]{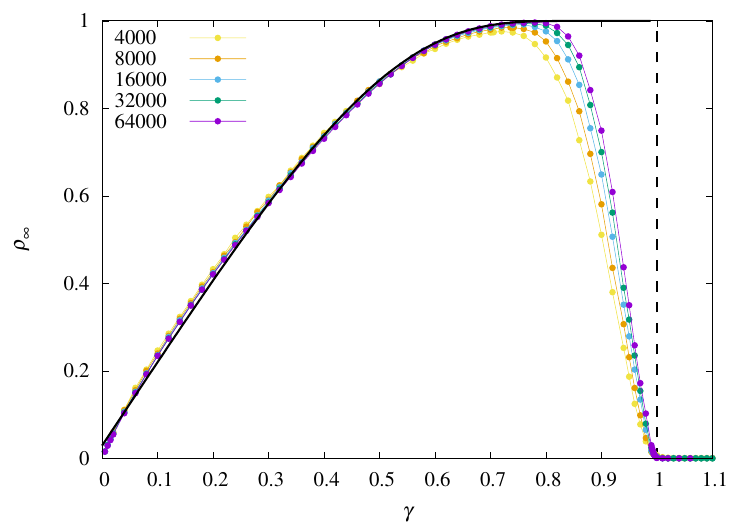}  
\caption{Mean fraction of agents in the largest group, $\rho_\infty$, for the quenched scenario as a function of the attachment exponent $\gamma$. The data points represent Monte Carlo simulation results for system sizes ranging from $N=4000$ to $N=64000$, as indicated. These quantities are measured in the final stationary state, which is reached when isolates can no longer join any existing group. The solid black curve represents a fit to the data for $N=64000$ in the region $\gamma < 0.8$, given by $\rho_\infty = 1 - a\exp[-b/(1-\gamma)]$ with $a = 6.96$ and $b=1.97$. The dashed vertical line indicates the critical value $\gamma_c^{(q)}=1$.
 }  
\label{fig:Qu1}  
\end{figure}

Figure \ref{fig:Qu1} displays the mean fraction of agents in the largest group, $\rho_\infty$, measured at equilibrium (frozen configuration). This order parameter is crucial as it reveals the conditions necessary for the formation of a consensus movement,   identified by $\rho_\infty \approx 1$. For a fixed system size $N$, $\rho_\infty$ increases monotonically with the attachment exponent $\gamma$, reaching a peak near $\gamma = 1$, before dropping abruptly to zero. In the thermodynamic limit, the results suggest a discontinuous transition from $\rho_\infty = 1$ to $\rho_\infty = 0$ at $\gamma = \gamma_c^{(q)} \approx  1$. The behavior in the regime $\gamma <  \gamma_c^{(q)}$ is well described by the function $\rho_\infty = 1 - a\exp[-b/(1-\gamma)]$, where $a$ and $b$ are fitting parameters. Conversely, for $\gamma \ge  \gamma_c^{(q)}$, we observe $\rho_\infty \to 0$ as $N$ increases. In this regime of ``demanding" agents (high thresholds), macroscopic groups fail to form.  Interestingly, this implies that for very large $N$, the formation of a massive protest requires an almost uniform distribution of thresholds ($\gamma \approx 1$), with only a marginal preference for smaller groups. The region near $\gamma \approx 1$ thus represents an optimal point for macroscopic organization, minimizing the density of independent entities.

\begin{figure}[th] 
\center
 \includegraphics[width=1\columnwidth]{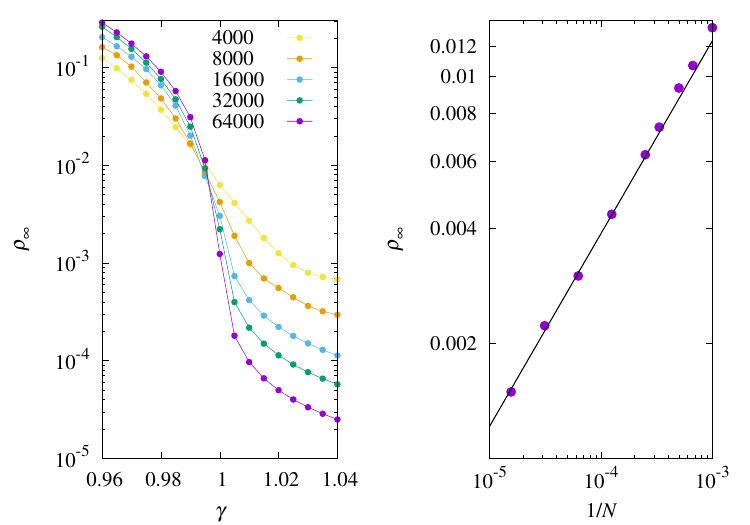}  
\caption{Left panel) Mean fraction of agents in the largest group, $\rho_\infty$, for the quenched scenario as a function of the attachment exponent $\gamma$ in the critical region. The intersection of curves for different system sizes $N$ occurs in the interval $(0.99, 1)$. (Right panel) Log-log plot showing the dependence of $\rho_\infty$ on $N$ at $\gamma_c^{(q)}=1$. The solid line represents the power-law fit $\rho_\infty = a N^{-b}$ with $a=0.407$ and $b=0.505$.
 }  
\label{fig:Qu2}  
\end{figure}

The existence of a discontinuity at $\gamma= \gamma_c^{(q)} $ is further supported by the results presented in Fig. \ref{fig:Qu2}, which zooms into the critical region to reveal the crossing of the $\rho_\infty$ vs. $\gamma$ curves for different system sizes $N$. The intersection points occur within the interval $\gamma \in (0.99,1)$ and shift towards $1$ as $N$ increases. The right panel of Fig. \ref{fig:Qu2} demonstrates that $\rho_\infty$ decays to zero following the power law $N^{-0.5}$ at the critical point $\gamma_c^{(q)}$. Due to the relatively fast freezing of the dynamics at this specific value of $\gamma$, we were able to average over $10^6$ independent simulations to generate the data for the right panel. This vanishing limit confirms that at $\gamma_c^{(q)}$ the largest group becomes sub-extensive ($S_{max} \sim N^{0.5}$), sharply contrasting with the macroscopic occupancy observed for $\gamma <\gamma_c^{(q)}$.

\begin{figure}[th] 
\center
 \includegraphics[width=\textwidth]{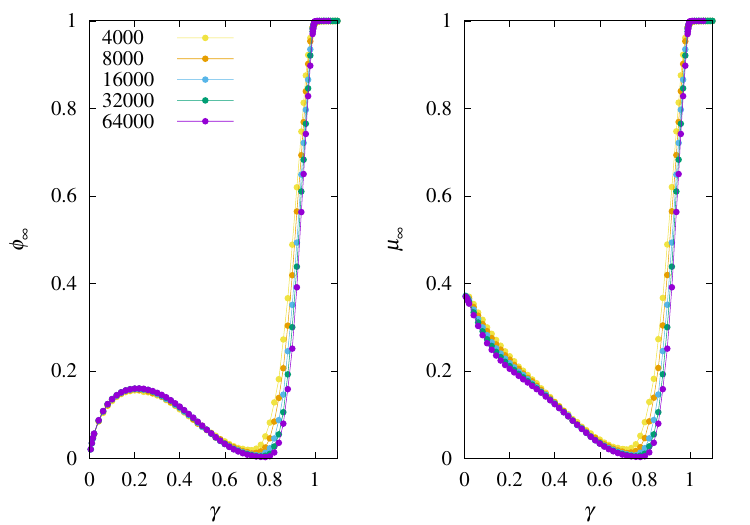}  
\caption{
Mean fraction of isolates $\phi_\infty$ (left panel) and mean density of groups $\mu_\infty$ (right panel) for the quenched scenario as a function of the attachment exponent $\gamma$. The data points represent Monte Carlo simulation results for system sizes ranging from $N=4000$ to $N=64000$, as indicated. These quantities are measured in the final stationary state, which is reached when the remaining isolates cannot join any of the existing groups.
 }  
\label{fig:Qu3}  
\end{figure}

A more comprehensive picture of the frozen configurations is obtained by analyzing the fraction of isolates $\phi_\infty$ and the density of groups $\mu_\infty$ as functions of $\gamma$, shown in Fig. \ref{fig:Qu3}. Although barely discernible at the scale of the figure, the curves for these quantities cross in the vicinity of $\gamma=1$ for different system sizes. At the point where the giant component $\rho_\infty$ reaches its maximum, both $\phi_\infty$ and $\mu_\infty$ attain their minima, approaching zero for large $N$. In the regime $\gamma \ge \gamma_c^{(q)}$, we observe $\phi_\infty \approx \mu_\infty \approx 1$. Consequently, the absorbing configurations are predominantly composed of isolates, with all other group sizes having a non-extensive representation.

In the limiting case $\gamma=0$, where agents have no preference regarding group size, the threshold is $T_i = 1$ for all $i$.  Thus,  the quenched and annealed scenarios become identical. In \ref{app:A}, we solve this case analytically in the thermodynamic limit, demonstrating that $\phi_\infty = 0$ (isolates are exhausted). The absorbing configurations are dominated by pairs, which account for $50\%$ of the groups. Triples and quadruples together comprise approximately $45.8\%$ of the groups. Furthermore, the probability that a randomly selected agent belongs to a group of size $k$ follows a shifted Poisson distribution, $w_k = e^{-1}/(k-2)!$ for $k \ge 2$. The analytical value for the total density of groups in this limit is $\mu_\infty = e^{-1} \approx 0.368$.

It is crucial to note that although the order parameter vanishes ($\rho_\infty = 0$) at both extremes of the spectrum ($\gamma=0$ and $\gamma \ge \gamma_c^{(q)}$), the physical nature of the absorbing configurations is fundamentally distinct. In the latter case ($\gamma \ge \gamma_c^{(q)}$), the system is in a ``frozen gas" state dominated by isolates due to the high demands of the agents. In contrast, for $\gamma=0$, the system is in a ``liquid-like" state dominated by small groups (pairs, triples), with isolates being completely absent.

\section{The Annealed Scenario}\label{sec:aa}

The previous analysis focused on the quenched scenario, where agent thresholds $T_i$ are generated at the beginning and remain fixed throughout the simulation. In this regime, the time scale associated with the random variables $T_i$ is effectively infinite compared to the aggregation process. However, the inherent heterogeneity of the thresholds renders the quenched  scenario analytically intractable. The annealed scenario, considered in this section, restores homogeneity by assuming that a new threshold is generated from the probability distribution $P(T_i)$ each time an agent attempts to select a group. Since the threshold distribution is identical for all agents, we can derive a closed set of rate equations  that are independent of the agent label $i$.

Let us consider the conditional expected values of the group populations $n_k(t + \delta t)$, given that the system is in state $\vec{n} = (n_1(t), \ldots, n_N(t))$, where $n_k(t)$ denotes the number of groups of size $k$ at time $t$. Recall that the total mass $\sum_{k} k n_k = N$ is conserved, whereas the total number of groups, $M = \sum_{k=1}^N n_k$, varies with time $t$. We assume here that the time interval $\delta t$ corresponds to a single elementary step during which a randomly chosen agent is selected and attempts to join a group if isolated, or remains in its current group otherwise.

Let us begin with the conditional expectation of the number of isolates, $n_1$. The number of isolates can only decrease in this dynamics (as disaggregation is excluded). An isolate disappears either when successfully joining an existing group or when merging with another isolate. We have
\begin{equation}
\mathbb{E} [ n_1 (t + \delta t) | \vec{n} ] = n_1 - \frac{n_1}{N(M-1)} \left[ \sum_{j=2}^N n_j \left(\frac{j}{N}\right)^\gamma + 2(n_1 - 1) \left ( \frac{1} {N} \right )^{\gamma} \right ] .
\end{equation}
The pre-factor $n_1/[N(M-1)]$ represents the joint probability of selecting an isolate to move ($n_1/N$) and this agent  choosing a specific group target among the $M-1$ available groups ($1/(M-1)$). The first term within the brackets represents the successful aggregation into groups of size $j \ge 2$, and the second term accounts for the joining of two isolates. The factor $2$ in the latter term reflects that, upon mutual aggregation ($1+1 \to 2$), both isolates leave the $n_1$ ensemble.

Next, we consider the conditional expectation of the number of  pairs,  $n_2$.  Pairs  are created by the merger of two isolates and destroyed upon receiving a third member.  We have
\begin{equation}
\mathbb{E} [ n_2 (t + \delta t) | \vec{n} ] = n_2 + \frac{n_1}{N(M-1)} \left[ (n_1 - 1)  \left ( \frac{1} {N} \right )^{\gamma}  - n_2 \left(\frac{2}{N}\right)^\gamma \right], 
\end{equation}
where the first term within the brackets represents the successful formation of a pair  resulting from the merging of two isolates ($1+1 \to 2$) and the second term represents the loss due to the growth into a  triple ($1+2 \to 3$). 

For groups with $k=3,4, \ldots, N$  agents,  the logic is uniform: a group grows if it acquires an isolate (transitioning from size $k-1$) and grows into the next size  (becoming $k+1$) upon receiving an isolate. There is no isolate self-interaction involved here, so we do not subtract 1 from the population counters $n$. We have 
\begin{equation}
\mathbb{E} [ n_k (t + \delta t) | \vec{n} ] = n_k + \frac{n_1}{N(M-1)} \left[ n_{k-1} \left(\frac{k-1}{N}\right)^\gamma - n_k \left(\frac{k}{N}\right)^\gamma \right], 
\end{equation}
where the first term within the brackets represents the successful formation of a group of size $k$  resulting from the merging of an isolate and a group of size $k-1$  and the second term represents the loss due to the growth into a  group of size $k+1$.

The next step is to introduce a mean-field approximation where the variables $n_k$ on the right-hand side of the evolution equations are replaced by their expected values $n_k \approx \mathbb{E}[n_k] = \langle n_k \rangle$ in the limit of large $N$. The continuous time limit
\begin{equation}
\frac{d \langle n_k \rangle }{dt} \approx \frac{ \langle n_k(t+\delta t) \rangle - \langle n_k(t) \rangle}{\delta t} 
\end{equation}
is obtained by setting the time step to $\delta t = 1/N$. It is worth noting that with this choice, one unit of continuous time $t$ corresponds to $N$ elementary time steps $\delta t$, which amounts to exactly one full Monte Carlo step.

To obtain a set of equations independent of the system size $N$ (the thermodynamic limit), we introduce the group densities $c_k = \langle n_k \rangle / N$ and the total group density $\mu = \langle M \rangle / N$. Furthermore, since the aggregation probability scales with $N^{-\gamma}$, we introduce a rescaled time variable $\tau = t N^{-\gamma}$.  In terms of these variables  and neglecting terms of order $1/N$ and higher,  we write the scale-invariant mean-field equations
\begin{eqnarray}
\frac{d c_1}{d\tau} & = & - \frac{c_1}{\mu} \left[ \sum_{j=2}^\infty c_j j^\gamma + 2 c_1 \right] \\
\frac{d c_k}{d\tau} & = & \frac{c_1}{\mu} \left[ c_{k-1} (k-1)^\gamma - c_k k^\gamma \right] \quad \text{for } k \ge 2.
\end{eqnarray}
These equations satisfy the normalized mass conservation $\sum_k  k c_k = 1$. The time evolution of the total group density $\mu = \sum_k c_k$ is given by
\begin{equation}\label{eq:mu}
\frac{d\mu}{d\tau} = - \frac{c_1}{\mu} \sum_{j=1}^\infty c_j j^\gamma.
\end{equation}
 Therefore,  both $c_1$ and  $\mu$ are monotonously decreasing functions of  time. 
From these equations it is evident that the  dynamics freezes only when $c_1 = 0$, i.e., when the reservoir of  isolates is empty.  This differs remarkably from the quenched scenario where freezing occurs when the group sizes are all below the thresholds of the remaining isolates.
The time required for the dynamics to  freeze scales as $t \sim N^\gamma$, reflecting the increasing difficulty for isolates to find suitable targets for not too small $\gamma$   as the system size grows.  

The limiting case $\gamma = 0$ is of particular interest as it admits a closed-form analytical solution, providing a rigorous benchmark for the numerical results. Physically, $\gamma = 0$ corresponds to a  situation where an isolate is equally likely to merge with any existing group regardless of its size.  As detailed in  \ref{app:A}, the mean-field equations can be solved exactly in this limit. The system follows a specific trajectory in the phase space $(c_1, \mu)$ given by
\begin{equation}
c_1(\mu) = \mu \left( 1 + \ln \mu \right).
\end{equation}
The dynamics proceed until the reservoir of isolates is depleted ($c_1=0$). This condition yields a final group density of $\mu_{\infty} = e^{-1} \approx 0.368$. Furthermore, the time evolution of the total group density is found to follow a double exponential decay
\begin{equation}
\mu(\tau) = \exp \left( e^{-\tau} - 1 \right).
\end{equation}
This result confirms that even in the absence of size preference, the system relaxes quickly towards a state with a finite group density.

Another integrable case arises when $\gamma = 1$. Physically, this corresponds to an aggregation probability that scales linearly with the target group size ($P(T \le k) \propto k$). This linearity allows us to utilize the mass conservation constraint to decouple the rate equations.
As derived in \ref{app:B}, the system evolves along a distinct trajectory in the phase space $(c_1, \mu)$ given by an exponential relation
\begin{equation}
c_1(\mu) = 2 e^{\mu-1} - 1.
\end{equation}
Unlike the $\gamma=0$ case, the preference for larger groups accelerates the depletion of isolates relative to the number of groups. The dynamics freeze when $c_1=0$, which corresponds to a final group density of $\mu_{\infty} = 1 - \ln 2 \approx 0.307$. This value is lower than in the $\gamma=0$ case ($\approx 0.368$), indicating a slightly higher degree of aggregation (fewer, larger groups) at equilibrium.

The temporal evolution is given implicitly by the integral relation
\begin{equation}
\int_{\mu(0)}^{\mu(\tau)} \frac{x}{2e^{x-1} - 1} dx = -\tau.
\end{equation}
While this integral cannot be expressed in terms of elementary functions (involving polylogarithms), it fully determines the kinetics of the system.

\begin{figure}[th] 
\center
 \includegraphics[width=1\columnwidth]{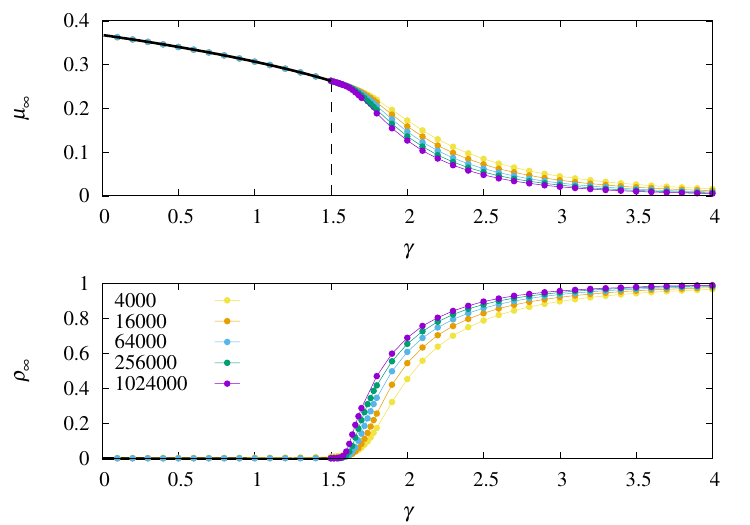}  
\caption{Mean density of groups $\mu_\infty$ (upper panel) and the mean fraction of agents in the largest group $\rho_\infty$ (lower panel) for the annealed scenario as a function of the attachment exponent $\gamma$. The data points represent Monte Carlo simulation results for system sizes ranging from $N=4000$ to $N=1024000$ as indicated. These quantities are measured in the final stationary state, characterized by the complete exhaustion of isolated agents ($n_1=0$). The solid black curve shows the analytical prediction from the mean-field rate equations. The dashed vertical line signals the critical  value, $\gamma_c^{(a)} \approx 1.499$.
 }  
\label{fig:An1}  
\end{figure}

\begin{figure}[th] 
\center
 \includegraphics[width=1\columnwidth]{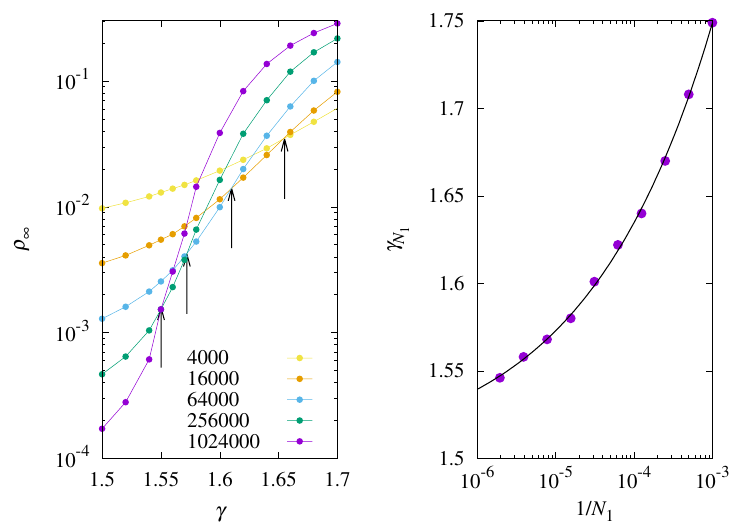}  
\caption{ (Left panel) Mean fraction of agents in the largest group $\rho_\infty$  for the annealed scenario as a function of the attachment exponent $\gamma$ in the critical region. The arrows show the intersection of curves for  consecutive system sizes $N$.  (Right panel) The intersection values $\gamma_{N_1}$, defined where the curves for $N_1$ and $N_2=2N_1$ cross, plotted as a function of the  inverse system size  $1/N_1$. The solid curve represents the scaling fit $\gamma_{N_1} = \gamma_c^{(a)} + a N_1^{-b}$, yielding parameters $a = 1.569 $, $b = 0.266$, and an asymptotic critical value $\gamma_c^{(a)} = 1.5$. Note the logarithmic scale on the $x$-axis.
 }  
\label{fig:An2}  
\end{figure}

For general $\gamma$ we can  solve the annealed mean-field equations numerically,  starting from an initial condition of pure isolates ($c_1 = 1$ and $c_k = 0$ for $k \ge 2$).  We find that for $\gamma \gtrsim 1.4$, the numerical  solution fails to converge.  This breakdown signals a violation  the mass conservation constraint $\sum k c_k = 1$,  caused   by a leakage of mass to infinity.   Consequently, the set of equations restricted to finite group sizes no longer accounts for the total mass of the system.  Crucially, the precise value of $\gamma$ at which this breakdown occurs is sensitive to implementation details, specifically the numerical threshold used to define non-zero densities.  For instance, enforcing a cutoff where $c_k$ is treated as zero if $c_k < 10^{-20}$ leads to a breakdown at $\gamma \approx 1.44$, whereas a looser threshold of $10^{-15}$ shifts this point to $\gamma \approx 1.52$. 
To resolve this ambiguity and better understand the transition, we rely on extensive finite-$N$ simulations, as presented in Figs.  \ref{fig:An1} and  \ref{fig:An2}. 

Simulations for the annealed scenario prove to be significantly more computationally demanding than for the quenched scenario. This stems from the fact that, for large $\gamma$ and $N$, the probability of a successful aggregation event becomes vanishingly small. To overcome the computational inefficiency of the standard rejection-based Monte Carlo method, we implemented a rejection-free algorithm. In the standard approach, an isolate randomly selects a target group and generates a threshold, often resulting in a rejected move if the target group size is insufficient. This leads to a vast number of null steps where the system state remains unchanged, causing prohibitively slow convergence towards the frozen state.

In the efficient rejection-free approach, we explicitly calculate the statistical weight (propensity) of all possible successful transitions available to an isolate in the current configuration. An active isolate has two distinct channels for aggregation: (i) nucleation, where it merges with another isolate to form a pair, with a weight proportional to the number of remaining isolates and the probability $P(T=1)$; and (ii) growth, where it joins an existing group of size $k \ge 2$, with a weight proportional to the acceptance probability $P(T \le k) \propto k^\gamma$. By constructing a cumulative distribution of these weights, we select the specific target for the next event in a single computational step. This method advances the system dynamics directly from one valid state to the next, eliminating wasted iterations and ensuring rapid convergence to the final state ($n_1=0$) regardless of the value of $\gamma$.

Figure \ref{fig:An1}  shows the mean group density $\mu_\infty$ and  the mean fraction of agents belonging to the largest group $\rho_\infty$,  both measured at the point where the dynamics freeze due to the exhaustion of isolates ($n_1=0$).  The results shown are averages over $10^4$ independent runs,   starting from the  all-isolates initial condition.  The physical mechanism underlying the non-existence of stationary mean-field solutions for large $\gamma$ is clarified by the behavior of the finite $N$ solutions: there is a critical value $\gamma_c^{(a)}$ beyond which $\rho_\infty $ becomes nonzero  in the thermodynamic limit $N \to \infty$.  This implies that for $\gamma > \gamma_c^{(a)}$, a macroscopic fraction of the total mass accumulates in a single massive group.  Consequently, the mass conservation constraint for finite clusters is violated ($\sum k c_k < 1$) due to the leakage of mass to the macroscopic scale, resulting in the breakdown of the  mean-field equations.

The discontinuous nature of the phase transition from $\rho_\infty=0$ to $\rho_\infty>0$ at $\gamma_c^{(a)}$ is revealed by the crossing of the $\rho_\infty$ vs. $\gamma$ curves for different system sizes $N$. While this feature is barely discernible in the lower panel of Fig. \ref{fig:An1}, it becomes prominent in the left panel of Fig. \ref{fig:An2} due to the use of a logarithmic scale and the focus on the critical region. The intersection points of curves for consecutive system sizes are identified by arrows. Let us denote by $\gamma_{N_1}$ the intersection point between the curves for $N_1$ and $N_2 = 2 N_1$. (Note that in Fig. \ref{fig:An1} and the left panel of Fig. \ref{fig:An2}, a ratio of $N_2 = 4 N_1$ was used for visualization purposes to avoid symbol cluttering.) The right panel of Fig. \ref{fig:An2} displays $\gamma_{N_1}$ for $N_1=1000 \cdot 2^j$ with $j=0, \ldots, 9$. The results indicate that the sequence $\gamma_{N_1}$ converges to $\gamma_c^{(a)} \approx 1.5$ following a power-law scaling of $N_1^{-0.266}$.

\begin{figure}[th] 
\center
 \includegraphics[width=1\columnwidth]{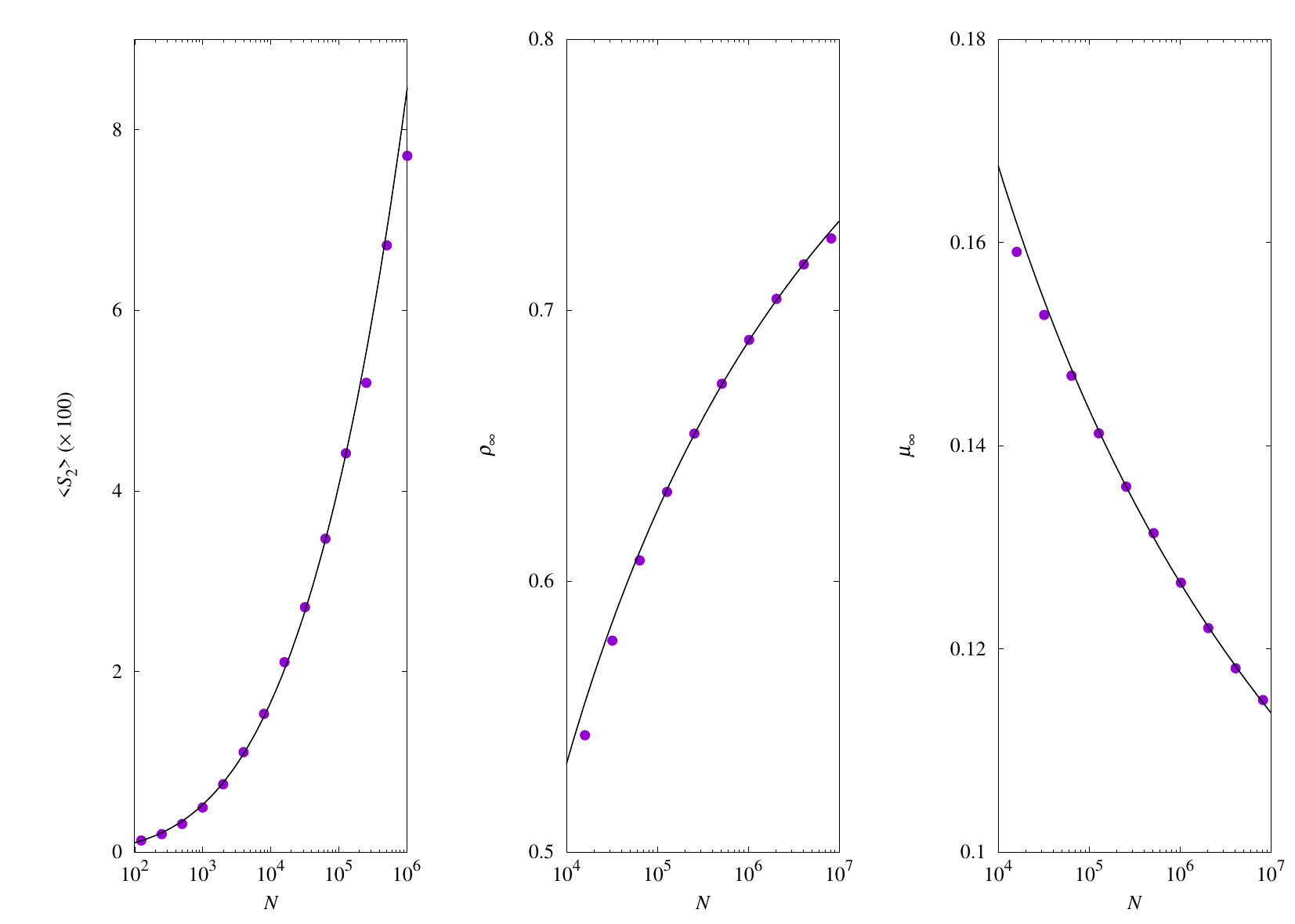}  
\caption{ (Left panel) Mean size of the second largest group, $\langle S_2 \rangle$, measured when the dynamics freezes, as a function of the system size $N$. The solid curve represents the fit $\langle S_2 \rangle = a \ln^b N$ with $a=0.023$ and $b=4$. (Middle panel) Mean fraction of agents in the largest group, $\rho_\infty$, as a function of $N$. The solid curve shows the fit $\rho_\infty = 1 - a/\ln^b N$ with $a=4.306$ and $b=1$. (Right panel) Mean density of groups, $\mu_\infty$, as a function of $N$. The solid curve corresponds to the fit $\mu_\infty = a/\ln^b N$ with $a=0.780$ and $b=0.693$. All data correspond to an attachment exponent of $\gamma = 2$  for the annealed scenario. 
 }  
\label{fig:An3}  
\end{figure}

Characterizing the nature of the condensed phase for $\gamma > \gamma_c^{(a)}$ in the thermodynamic limit is particularly challenging due to the extremely slow convergence with increasing $N$, as depicted in Fig. \ref{fig:An1}. The first issue we address is whether the violation of mass conservation stems from the emergence of a single massive group (consensus) or the coexistence of multiple macroscopic groups (factionalism). To distinguish between these scenarios, the left panel of Fig. \ref{fig:An3} presents the mean size of the second largest group, $\langle S_2 \rangle$, measured at the freezing point, as a function of the system size $N$ for $\gamma = 2$ (deep within the condensed phase). Due to large fluctuations in $S_2$, the data points in this panel represent averages over $10^5$ independent realizations. We find that $\langle S_2 \rangle$ grows sub-linearly with $N$; specifically, the data is well described by $\langle S_2 \rangle \propto \ln^4 N$. This confirms that the condensed phase is characterized by a single macroscopic cluster capturing a finite fraction of the system's mass. This ``winner-takes-all" scenario is consistent with the preferential attachment mechanism becoming super-linear for $\gamma >  \gamma_c^{(a)}$, which amplifies small initial fluctuations and allows the leading group to outgrow and suppress all competitors.

The next critical issue is whether the largest group encompasses the entire population or coexists with a macroscopic fraction of the total mass distributed among microscopic groups in the thermodynamic limit. The results shown in the middle panel of Fig. \ref{fig:An3} suggest that $\rho_\infty \to 1$ for $\gamma = 2$ as $N \to \infty$. This implies that the largest group absorbs nearly the entire population, leaving behind only a non-extensive residue of small groups whose sizes scale, at most, as $\ln^4 N$. In this scenario, the density of groups must vanish ($\mu_\infty \to 0$) in the thermodynamic limit, a conclusion corroborated by the results shown in the right panel of Fig. \ref{fig:An3}. We note that these conclusions regarding the asymptotic values of $\rho_\infty$ and $\mu_\infty$ rely on the assumption that the convergence to the asymptotic limit is governed by a power of the logarithm of the system size, $1/\ln^b N$.  This logarithmic convergence is likely a consequence of the persistent competition with the second largest group; since its size also increases with the system size ($\langle S_2 \rangle \sim \ln^4 N$), the absorption of the remaining mass into the giant component is retarded, slowing down the convergence of $\rho_\infty$ to unity.

\section{Discussion}\label{sec:disc}

The approach proposed in this work generalizes the seminal threshold model of collective behavior by moving beyond the binary paradigm of simple activation. While standard threshold models typically describe an individual's decision to participate in a single collective action \cite{Granovetter_1978}, our framework addresses the more complex scenario of competitive aggregation. In reality, individuals often face a ``marketplace" of social movements: they may choose among various simultaneous protests, political factions, or social causes, but their capacity for active engagement is often limited to a single commitment \cite{McCarthy_1977}. Furthermore, we assume a regime of high peer pressure or high switching costs, such that once  individuals join a group, they do not leave. This irreversible competition is observable in various real-world contexts, such as the initial stages of political uprisings where multiple factions compete for the loyalty of the populace \cite{Bakke_2012}, or in the emergence of dominant online communities where network effects create strong lock-in mechanisms \cite{Arthur_1989}.

A central finding of this study is that the system's macroscopic behavior is determined not just by the average agent behavior, but by the specific nature of the disorder---the inherent heterogeneity in individual thresholds. We show that the way this disorder is implemented, whether as a fixed attribute of the agents (quenched) or as a fluctuating environmental variable (annealed), fundamentally alters the path to consensus. This distinction is critical because it suggests that two populations with identical statistical distributions of  ``stubbornness" can reach diametrically opposed social outcomes based solely on the temporal stability of those individual predispositions. We have argued that the quenched scenario is the more faithful representation of physical social systems where thresholds reflect stable psychological traits or socio-economic risks. However, the annealed scenario---where thresholds are drawn anew at each interaction---provides more than just mathematical convenience. It represents a realistic description of modern digital environments characterized by high information churn. In social media platforms, for instance, an individual's threshold for engagement may fluctuate based on current cognitive load, the temporary emotional framing of a post, or the algorithmic sequence of their feed \cite{Lorenz_2019, Weng_2012}. Similar annealed dynamics are observed in financial markets, where an agent's threshold to trade is reset by every new price tick \cite{Cont_2001}, or in emergency evacuations, where panic and immediate environmental cues override pre-existing dispositions \cite{Helbing_2000}.

The results presented in this work reveal a striking contrast between the quenched and annealed scenarios. Although both frameworks share an identical underlying threshold probability distribution governed by the exponent $\gamma$, their dynamical consequences are fundamentally divergent. In the quenched scenario, $\gamma$ dictates the static landscape of individual resistance: it defines who will participate and who will remain isolated. When $\gamma \geq \gamma_c^{(q)} \approx 1$, the abundance of  stubborn  agents triggers a generic coordination failure, effectively pinning the system in a fragmented state where the relative size of the largest group,  $\rho_\infty$,  vanishes. Conversely, for $\gamma < \gamma_c^{(q)}$, a sufficient density of flexible early-adopters facilitates the emergence of macroscopic collective action.  In the thermodynamic limit, as $\gamma$  increases from zero to $\gamma_c^{(q)}$,  $\rho_\infty$ increases monotonically toward unity.  At $\gamma= \gamma_c^{(q)}$,  $\rho_\infty$ jumps to zero. 
In finite systems, we observe that $\rho_\infty$ peaks as $\gamma$ approaches the critical threshold  $\gamma_c^{(q)}$ from below, followed by an abrupt collapse to zero.

In the annealed scenario, the constant renewal of thresholds eliminates permanent agent stubbornness, and $\gamma$ effectively plays the role of a dynamic attachment exponent.  For $\gamma > \gamma_c^{(a)} \approx 1.5$,  this attachment preference becomes super-linear, triggering a ``winner-takes-all" dynamic. A group that acquires a slight size advantage becomes disproportionately attractive, growing explosively and suppressing the nucleation of competitors. This leads to a consensus state where a single massive group encompasses the entire population ($\rho_\infty \to 1$). On the other hand, for $\gamma < \gamma_c^{(a)} $,  the advantage of being large is weak and the rate of nucleation of new pairs competes effectively with the growth of existing groups, resulting in a state pulverized into many small groups with no macroscopic aggregation ($\rho_\infty \to 0$). Thus, a high $\gamma$ prevents consensus in the quenched case by blocking initiation, whereas it enforces consensus in the annealed case by suppressing competition.

From a sociophysical perspective, these findings suggest that even when multiple competing alternatives exist, the formation of a unified, massive movement is acutely sensitive to the distribution of individual risk thresholds. The discontinuous transitions at $\gamma_c^{(q)}$ and $\gamma_c^{(a)}$ function as critical social tipping points that dictate the viability of collective action. In the quenched case (Fig. \ref{fig:Qu1}), the strategy for triggering a movement that encompasses the entire population appears remarkably precarious. A population dominated by too many low-threshold instigators—individuals who catalyze initial growth with minimal social reinforcement—paradoxically results in  pulverization,  where participants are scattered across numerous disconnected factions. While a massive movement is only achieved as $\gamma$ approaches the critical threshold $\gamma_c^{(q)}$, this state is inherently fragile; a marginal increase in $\gamma$ beyond this optimum leads to a total collapse of coordination. Conversely, an authority might seek to prevent such a tipping point by strategically raising the effective $\gamma$—for instance, through targeted deterrence or by sowing skepticism to disenfranchise these initial activists. However, this is a risky gamble for the state: an attempt to suppress a movement by artificially inflating $\gamma$ may inadvertently tune the population to the optimal threshold, unintentionally facilitating the very macroscopic mobilization the authority sought to prevent.   

The deterrence perspective reveals a more manageable, albeit counter-in\-tui\-ti\-ve, logic in the annealed scenario (lower panel of Fig. \ref{fig:An1}). Here, the authority's most effective strategy to maintain stability is not necessarily to suppress participation through high $\gamma$, but rather to facilitate a  ``divide-and-conquer" outcome by decreasing the effective threshold. By prompting individuals to join a multitude of distinct, small-scale movements, the state avoids the threat of a singular macroscopic mobilization. In this regime, the population remains highly active but functionally neutralized by extreme fragmentation---effectively trading the risk of a revolutionary surge for the controlled chaos of a pulverized social landscape.

Finally, we identify several promising directions for future research. A natural extension of this framework is to relax the assumption of irreversibility by allowing disaggregation, where individuals may leave a group with a certain probability, as done in the casual groups framework \cite{Fontanari_2023}. This would transform the absorbing frozen states into dynamic steady states, allowing for a detailed study of the fluctuations and the stability of the giant component. Furthermore, introducing a non-trivial topology for the interactions \cite{Wiedermann_2020}, rather than the mean-field approach assumed here, could reveal how social network structures influence the competition between local factions and global consensus.

\section*{Acknowledgments}
JFF is partially supported by  Conselho Nacional de Desenvolvimento Cient\'{\i}fico e Tecnol\'ogico  grant number 305620/2021-5.  BYSI is supported by  Fun\-da\-\c{c}\~ao de Amparo \`a Pesquisa do Estado de S\~ao Paulo 
(FAPESP) grant number 25/06044-7.





\appendix

\renewcommand{\theequation}{A.\arabic{equation}}
\setcounter{equation}{0}
\renewcommand{\thefigure}{A\arabic{figure}}
\setcounter{figure}{0}

\section{Annealed solution for $\gamma = 0$ }\label{app:A}

In the limit $\gamma = 0$, the term $j^\gamma$ becomes unity for all $j$. Consequently, the summation terms in the rate equations simplify to the zeroth moment of the distribution, i.e., $\sum_{j=1}^\infty c_j = \mu$. The rate equation for the total cluster density simplifies to
\begin{equation}
\frac{d\mu}{d\tau} = - \frac{c_1}{\mu} \sum_{j=1}^\infty c_j = - c_1.
\label{eq:app_dmu}
\end{equation}
Similarly, the evolution equation for the isolate density becomes
\begin{equation}
\frac{dc_1}{d\tau} = - \frac{c_1}{\mu} \left[ \sum_{j=2}^\infty c_j + 2 c_1 \right] = - \frac{c_1}{\mu} \left[ (\mu - c_1) + 2 c_1 \right] = - \frac{c_1}{\mu} (\mu + c_1).
\label{eq:app_dc1}
\end{equation}
Dividing Eq. (\ref{eq:app_dc1}) by Eq. (\ref{eq:app_dmu}), we eliminate time and obtain a linear first-order differential equation for the phase space trajectory
\begin{equation}
\frac{dc_1}{d\mu} = 1 + \frac{c_1}{\mu}.
\end{equation}
Using the integrating factor $1/\mu$ and applying the initial conditions $c_1(0) = \mu(0) = 1$ (all agents are isolates), integration yields
\begin{equation}
c_1(\mu) = \mu (1 + \ln \mu).
\label{eq:c1_mu}
\end{equation}
Substituting $c_1$ back into Eq. (\ref{eq:app_dmu}), we obtain
\begin{equation}
\int \frac{d\mu}{\mu (1 + \ln \mu)} = - \int d\tau.
\end{equation}
Performing the integration gives $\ln(1 + \ln \mu) = -\tau$, which leads directly to the solution for the total density
\begin{equation}
\mu(\tau) = \exp(e^{-\tau} - 1).
\label{eq:mu_tau}
\end{equation}
Using this result in Eq. (\ref{eq:c1_mu}), we find the explicit time dependence of the isolate density
\begin{equation}
c_1(\tau) = \mu(\tau) (1 + e^{-\tau} - 1) = \mu(\tau) e^{-\tau}.
\end{equation}

We can further generalize this to find the density $c_k(\tau)$ for any group size $k$. For $k \ge 2$, the rate equation is
\begin{equation}
\frac{dc_k}{d\tau} = \frac{c_1}{\mu} (c_{k-1} - c_k).
\end{equation}
Noting that $c_1/\mu = e^{-\tau}$, let us introduce a transformed time variable $T(\tau) = 1 - e^{-\tau}$, such that $dT = e^{-\tau} d\tau$. The equation becomes
\begin{equation}
\frac{dc_k}{dT} + c_k = c_{k-1},
\end{equation}
with initial conditions $c_k(0) = 0$ for $k \ge 2$. Using the integrating factor $e^T$ and the solution for $c_1(\tau) = e^{-T} (1-T)$, we can solve this hierarchy recursively. For $k=2$
\begin{equation}
\frac{d}{dT}(e^T c_2) = e^T c_1 = 1 - T \implies c_2(T) = e^{-T} \left( T - \frac{T^2}{2} \right).
\end{equation}
By induction, the general solution is
\begin{equation}
c_k(\tau) = \mu(\tau) \left[ \frac{(1 - e^{-\tau})^{k-1}}{(k-1)!} - \frac{(1 - e^{-\tau})^k}{k!} \right].
\end{equation}
This expression completely describes the cluster size distribution at any time. In the asymptotic limit $\tau \to \infty$, we have $e^{-\tau} \to 0$, yielding the frozen distribution
\begin{equation}
c_k(\infty) = e^{-1} \left[ \frac{1}{(k-1)!} - \frac{1}{k!} \right] = e^{-1} \frac{k-1}{k!}.
\label{eq:ck_inf}
\end{equation}
From Eq. (\ref{eq:ck_inf}), we can define the normalized probability $p_k$ that a randomly chosen group has size $k$
\begin{equation}
p_k = \frac{c_k(\infty)}{\mu(\infty)} = \frac{e^{-1} (k-1)/k!}{e^{-1}} = \frac{k-1}{k!} \quad \text{for } k \ge 2.
\end{equation}
The mode of this distribution occurs at $k=2$ ($p_2 = 1/2$), followed by $p_3 = 1/3$ and $p_4 = 1/8$, confirming a dominance of small clusters.

Interestingly, the mass-weighted distribution $w_k = k c_k(\infty)$, which represents the probability that a randomly selected agent belongs to a group of size $k$, takes the form of a shifted Poisson distribution
\begin{equation}
w_k = k c_k(\infty) = e^{-1} \frac{k(k-1)}{k!} = \frac{e^{-1}}{(k-2)!} \quad \text{for } k \ge 2.
\end{equation}
This implies that the size of the group to which an agent belongs, shifted by 2, follows a standard Poisson distribution with parameter $\lambda=1$ (i.e., $S-2 \sim \text{Poisson}(1)$).

\section{Annealed solution for $\gamma = 1$ }\label{app:B}
\renewcommand{\theequation}{B.\arabic{equation}}
\setcounter{equation}{0}
\setcounter{figure}{0}

For $\gamma = 1$, the aggregation kernel is linear ($j^\gamma = j$). Crucially, the summation term appearing in the rate equations corresponds to the first moment of the distribution, which is conserved by definition: $\sum_{j=1}^\infty c_j j = 1$ (normalized mass density).

Substituting this into the rate equation for $\mu$ (Eq. \ref{eq:mu})
\begin{equation}
\frac{d\mu}{d\tau} = - \frac{c_1}{\mu} \sum_{j=1}^\infty c_j j = - \frac{c_1}{\mu} \cdot 1 = - \frac{c_1}{\mu}.
\label{eq:app_dmu_g1}
\end{equation}
For the isolate density $c_1$, the equation becomes
\begin{equation}
\frac{dc_1}{d\tau} = - \frac{c_1}{\mu} \left[ \sum_{j=2}^\infty c_j j + 2 c_1 \right].
\end{equation}
Using the mass conservation $\sum_{j=2}^\infty c_j j = 1 - c_1$, we simplify the term in brackets
\begin{equation}
\frac{dc_1}{d\tau} = - \frac{c_1}{\mu} \left[ (1 - c_1) + 2 c_1 \right] = - \frac{c_1}{\mu} (1 + c_1).
\label{eq:app_dc1_g1}
\end{equation}
Dividing Eq. (\ref{eq:app_dc1_g1}) by Eq. (\ref{eq:app_dmu_g1}) yields the trajectory equation
\begin{equation}
\frac{dc_1}{d\mu} = \frac{-(c_1/\mu)(1+c_1)}{-c_1/\mu} = 1 + c_1.
\end{equation}
This is a linear ODE whose  general solution is $c_1(\mu) = A e^\mu - 1$. Applying the initial conditions $c_1(0) = \mu(0) = 1$
\begin{equation}
1 = A e^1 - 1 \implies A = \frac{2}{e}.
\end{equation}
Thus, $c_1(\mu) = 2e^{\mu-1} - 1$.

Finally, to find the time dependence, we substitute $c_1$ back into Eq. (\ref{eq:app_dmu_g1}
\begin{equation}
\frac{d\mu}{d\tau} = - \frac{1}{\mu} \left( 2e^{\mu-1} - 1 \right).
\end{equation}
Separating variables leads to the integral form presented in the main text.

It is also possible to obtain closed-form solutions for the cluster densities $c_k$ in the asymptotic limit $\tau \to \infty$. By performing the change of variables $x = e^{\mu-1}$, the rate equations for $k \ge 2$ can be rewritten as
\begin{equation}
x \frac{dc_k}{dx} - k c_k = -(k-1) c_{k-1}.
\end{equation}
Using $c_1(x) = 2x-1$, we can solve this hierarchy recursively. The solutions are polynomials in $x$
\begin{align}
c_2(x) &= -\frac{3}{2}x^2 + 2x - \frac{1}{2} \\
c_3(x) &= \frac{4}{3}x^3 - 3x^2 + 2x - \frac{1}{3} \\
c_4(x) &= -\frac{5}{4}x^4 + 4x^3 - \frac{9}{2}x^2 + 2x - \frac{1}{4} \\
c_5(x) &= \frac{6}{5}x^5 - 5x^4 + 8x^3 - 6x^2 + 2x - \frac{1}{5}
\end{align}
In the asymptotic limit, $c_1 =0$ implies $x=1/2$. Evaluating the polynomials at this point yields the final frozen densities
\begin{equation}
c_k(1/2) = \frac{k-1}{k 2^k}.
\end{equation}
For example, $c_2(1/2) = 1/8$, $c_3(1/2) = 1/12$, $c_4(1/2) = 3/64$, and $c_5(1/2) = 1/40$. 
 We can immediately verify the consistency of this result by checking that the  mass conservation condition $\sum_{k=1}^\infty k c_k(1/2) = 1$ is satisfied.

\end{document}